# Application of EOS-ELM with Binary Jaya-Based Feature Selection to Real-Time Transient Stability Assessment Using PMU Data

Yang Li, *Member, IEEE*, Zhen Yang

*Abstract*— Recent researches show that pattern recognition-based transient stability assessment (PRTSA) is a promising approach for predicting the transient stability status of power systems. However, many of the current well-known PRTSA methods suffer from excessive training time and complex tuning of parameters, resulting in inefficiency for real-time implementation and lacking the on-line model updating ability. In this paper, a novel PRTSA approach based on an ensemble of OS-ELM (EOS-ELM) with binary Jaya (BinJaya)-based feature selection is proposed with the use of PMU data. After briefly describing the principles of OS-ELM, an EOS-ELM-based PRTSA model is built to predict the post-fault transient stability status of power systems in real time by integrating OS-ELM and an online boosting algorithm respectively as a weak classifier and an ensemble learning algorithm. Furthermore, a BinJaya-based feature selection approach is put forward for selecting an optimal feature subset from the entire feature space constituted by a group of system-level classification features extracted from PMU data. The application results on the IEEE 39-bus system and a real provincial system show that the proposal has superior computation speed and prediction accuracy than other state-of-the-art sequential learning algorithms. In addition, without sacrificing the classification performance, the dimension of the input space has been reduced to about one-third of its initial value.

## I. INTRODUCTION

TRANSIENT stability assessment (TSA) of power systems has always been regarded as a primary task to guarantee the system safe and stable operation [1]. With problems arising from electricity market reforms, the increasing application of power electronic devices and the gird integration of large-scale renewable resources, the dynamic behaviors of modern power systems are becoming more and more complex [2, 3], and the consequences resulted from transient instability are growing increasingly serious therewith [4-6]. Therefore, it is an urgent need for developing a well-calibrated TSA approach to make a fast and accurate determination of the transient stability status of post-fault power systems.

Transient stability is the ability of synchronous machines to maintain synchronism when subjected to a severe disturbance.

It is dependent not only on the initial operating state of the pre-fault system, but also on the disturbance's severity [7]. Since transient stability is a very fast phenomenon that requires a corrective control action within short period of time (< 1 s) [1, 8], fast detection of instability is essential.

In literature, the existing TSA methods can be divided into four basic classes: time-domain (T-D) simulations [9], direct methods (e.g. transient energy function (TEF) methods [10, 11] and the extended equal-area criterion (EEAC) [12]), and Lyapunov exponents (LEs) methods [13, 14] and pattern recognition-based TSA (PRTSA) methods. The T-D simulation is most straightforward approach with high-accuracy calculation results, but it is time-consuming and the results are strongly dependent on the accuracy of the system model and parameters [1]. The direct methods have fast calculation speed and are able to provide transient stability margins, but there are still several open problems existing to determine the specific TEF or coherent generator groups under a certain disturbance when using this approach in practical power systems with complex models [10, 12]. The LEs method proved that the transient stability can be determined by identifying the sign of the system's maximal Lyapunov exponent (MLE) [13]. However, the LEs methods needs several seconds to calculate MLE due to the limitations of observed time window length [13, 14].

In recent years, PRTSA has proved to be potential in the area of on-line dynamic security analysis by applying of machine learning techniques, such as artificial neural networks [15-17], support vector machine [18-20], decision trees [21, 22], and core vector machine [23], for solving protection and control problems of power systems. Form the viewpoint of PRTSA, the TSA problem can be viewed as a pattern recognition task, and the transient stability can be assessed by mapping relationships between input features extracting from the system operational parameters and final post-fault stability status [24-26]. Meanwhile, the matured applications of phasor measurement units (PMU)-based wide area measurement system (WAMS) have made it become a reality to acquire the real-time synchronized measurements, and this brings new ideas and opportunities for implementing an advanced wide-area protection and control system [27, 28].

Unfortunately, many of the well-known PRTSA methods suffer from excessive training time and complex tuning of

Yang Li (corresponding author) and Zhen Yang are with the School of Electrical Engineering, Northeast Electric Power University (NEEPU), Jilin 132012, P.R. China (e-mail: liyang@neepu.edu.cn).

parameters [15, 16, 18-20, 25], resulting in inefficiency for real-time implementation and lacking the on-line model updating ability. The traditional framework of in such approaches is the application mode of "offline training-online matching" [15-20]. When a trained model is unsatisfactory for some special samples in on-line application, the running model has to be terminated and offline retrained again [21]. However, in a real operating environment, training samples cannot cover all of the operating modes of time-varying modern complex power systems for sure. This will inevitably lead to the deteriorated applicability of the trained model via off-line training when used online [17], because training samples generated by offline simulations might not be able to represent the current modes.

ELM proposed by Huang is a new machine learning approach for single hidden layer feed forward networks [29], and it has been successfully applied in many engineering applications [30-33]. As an extension of ELM, OS-ELM can learn data one-by-one or chunk-by-chunk and discard the data for which the training has already been done [34, 35]. Ensemble learning is an attracting machine learning approach [22, 36-38], which learns knowledge by using a set of learning machines and comprehensively ensembles various learning results to obtain better generalization ability than individual learning machines. Besides classifier design, it is well-known that, for PRTSA, feature selection is of paramount importance [24, 25]. The idea is that feature selection will improve the classifier performance and provide a faster classification, leading to comparable or even best generalization ability than using all features [26].

In this paper, a novel PRTSA approach based on an ensemble of OS-ELM (EOS-ELM) with binary Jaya (BinJaya)-based feature selection is proposed with the use of PMU data. Use of ELM and OS-ELM for TSA have been previously studied in [31] and [35], respectively. The aim of that method in [31] is to trigger preventive control as a precaution for a set of contingencies, and its inputs are extracted from pre-fault steady-state information. Different from the TSA model in [31], a PRTSA model is built to predict the post-fault transient stability status of a power system in real time with consideration of post-fault dynamic-state information in this work. In [35], OS-ELM is introduce into TSA to overcome the inefficiency of online model updating existing in many of current models due to trivial parameter tuning. However, recent research suggests that original OS-ELM has the drawback of weak stability in different trials [36]. In addition, the used input features in [35] may not always be the 'best' ones for different cases since they are primarily selected through simulation analysis without a feature selection procedure. To overcome these issues, we utilize EOS-ELM to further improve the stability and accuracy of original OS-ELM. Besides, a BinJaya-based feature selection approach is put forward for selecting an optimal feature subset from the entire feature space constituted by a group of system-level classification features extracted from PMU data.

The remainder of this paper is structured as follows: Section II gives a brief introduction of the basic principles of OS-ELM. Next, a detailed description of the proposal using EOS-ELM is put forward in Section III. Section IV provides a novel BinJaya-based feature selection approach for EOS-ELM, with Section V examining the proposal on the IEEE 39-bus system and a real provincial power system in China. And finally, the conclusions are drawn from the simulation results.

## II. PRINCIPLES OF OS-ELM

As a kind of typical batch learning algorithms, ELM is hard to satisfy the demand for online updating of TSA models [35]. In such cases, OS-ELM is no doubt a better choice due to no retraining from scratch whenever a new sample arrives.

OS-ELM can be summarized in the following two steps [34]:

*Step 1:* **Initialization phase**.

For the given training set $D$, a small chunk of initial training data $D_0 = \left\{ (\mathbf{x}_i, \mathbf{t}_i) \middle| \mathbf{x}_i \in \mathbf{R}^n, \mathbf{t}_i \in \mathbf{R}^m, i = 1, \cdots, N_0 \right\}$ is chosen from $D$ to initialize the learning, $N_0 \geq L$. Here, $N_0$ is the number of samples in $D_0$, and $L$ is the number of hidden layer nodes.

(a) The hidden node parameters $(\mathbf{a}_i, b_i)$, $i = 1, \cdots, L$, are randomly generated. Here, $\mathbf{a}_i$ is the input weights vector, and $b_i$ is the bias of the $i$-th hidden node.

(b) The initial hidden layer output matrix $\mathbf{H}_0$ is obtained as

$$\mathbf{H}_0 = \begin{bmatrix} G(a_1, b_1, x_1) & \cdots & G(a_L, b_L, x_1) \\ \vdots & \cdots & \vdots \\ G(a_1, b_1, x_{N_0}) & \cdots & G(a_L, b_L, x_{N_0}) \end{bmatrix}_{N_0 \times L} \quad (1)$$

where $G(\cdot)$ denotes an activation function.

(c) The initial output weight $\boldsymbol{\beta}^{(0)}$ is estimated according to

$$\boldsymbol{\beta}^{(0)} = \mathbf{P}_0 \mathbf{H}_0^T \mathbf{T}_0 \quad (2)$$

where $\mathbf{P}_0 = (\mathbf{H}_0^T \mathbf{T}_0)^{-1}$, and $\mathbf{T}_0 = [\mathbf{t}_1, \cdots, \mathbf{t}_{N_0}]^T$.

(d) Set $k = 0$, where $k$ is the number of chunks.

*Step 2:* **Sequential learning phase**.

(a) Present the $(k+1)$th chunk of new observations:

$$D_{k+1} = \left\{ (\mathbf{x}_i, \mathbf{t}_i) \right\}_{l = (\sum_{j=0}^{k} N_j) + 1}^{\sum_{j=0}^{k+1} N_j} \quad (3)$$

where $N_{k+1}$ denotes the number of observations in the $(k+1)$th chunk.

(b) The partial hidden layer output matrix $\mathbf{H}_{k+1}$ is calculated for the $(k+1)$th chunk of data $D_{k+1}$:

$$\mathbf{H}_{k+1} = \begin{bmatrix} G\left(\mathbf{a}_1, b_1, \mathbf{x}_{(\sum_{j=0}^{k} N_j)+1}\right) & \cdots & G\left(\mathbf{a}_L, b_L, \mathbf{x}_{(\sum_{j=0}^{k} N_j)+1}\right) \\ \vdots & \cdots & \vdots \\ G\left(\mathbf{a}_1, b_1, \mathbf{x}_{\sum_{j=0}^{k+1} N_j}\right) & \cdots & G\left(\mathbf{a}_L, b_L, \mathbf{x}_{\sum_{j=0}^{k+1} N_j}\right) \end{bmatrix}_{N_{k+1} \times L} \quad (4)$$

Set $\mathbf{T}_{k+1} = \left[ \mathbf{t}_{(\sum_{j=0}^{k} N_j)+1}, \cdots, \mathbf{t}_{\sum_{j=0}^{k+1} N_j} \right]^T$.

(c) The output weight $\boldsymbol{\beta}^{(k+1)}$ is determined according to the following equations.

$$\mathbf{P}_{k+1} = \mathbf{P}_k - \mathbf{P}_k \mathbf{H}_{k+1}^T \left( \mathbf{I} + \mathbf{H}_{k+1} \mathbf{P}_k \mathbf{H}_{k+1}^T \right)^{-1} \mathbf{H}_{k+1} \mathbf{P}_k \quad (5)$$

$$\boldsymbol{\beta}^{(k+1)} = \boldsymbol{\beta}^{(k)} + \mathbf{P}_{k+1} \mathbf{H}_{k+1}^T \left( \mathbf{T}_{k+1} - \mathbf{H}_{k+1} \boldsymbol{\beta}^{(k)} \right) \quad (6)$$

(d) Set $k = k + 1$. Go to *step 2*(a).

When the training samples are received in the mode of one-by-one, $N_{k+1} \equiv 1$, equation (5) and (6) can be respectively the following simple format:

$$\mathbf{P}_{k+1} = \mathbf{P}_k - \frac{\mathbf{P}_k \mathbf{h}\left(\mathbf{x}_{k+1}\right) \mathbf{h}^T\left(\mathbf{x}_{k+1}\right) \mathbf{P}_k}{1 + \mathbf{h}^T\left(\mathbf{x}_{k+1}\right) \mathbf{P}_k \mathbf{h}\left(\mathbf{x}_{k+1}\right)} \quad (7)$$

$$\boldsymbol{\beta}^{(k+1)} = \mathbf{P}_{k+1} \mathbf{h}\left(\mathbf{x}_{k+1}\right)\left(\mathbf{t}_{k+1}^T - \mathbf{h}^T\left(\mathbf{x}_{k+1}\right)\boldsymbol{\beta}^{(k)}\right) + \boldsymbol{\beta}^{(k)} \quad (8)$$

where $\mathbf{h}\left(\mathbf{x}_{k+1}\right) = \left[G\left(\mathbf{a}_1, b_1, \mathbf{x}_{k+1}\right) \cdots G\left(\mathbf{a}_L, b_L, \mathbf{x}_{k+1}\right)\right]$.

## III. TSA BASED ON EOS-ELM

### A. EOS-ELM

Considering the output errors of OS-ELM are volatile due to random assignment of the hidden-node parameters, ensemble learning is introduced to improve the classification ability. Here, OS-ELM is employed as a weak classifier, and an online boosting algorithm is used as an ensemble learning algorithm.

#### 1) Concepts of online boosting

Supposed that a set of $M$ weak classifiers are given with the hypothesis $H^{weak} = \{h_1^{weak}, \cdots, h_M^{weak}\}$, a selector is employed to select one of those classifiers.

$$h^{sel}(x) = h_m^{weak}(x) \quad (9)$$

where $m$ is determined in term of an optimization criterion [39]. Factually, the estimated error $e_i$ of every weak classifier $h_i^{weak} \in H^{weak}$ is employed in the process. Specifically speaking, the corresponding index of the weak classifier with the lowest estimated error $e_i$ is chosen as the parameter $m$, which is defined as

$$m = \arg\min_i e_i \quad (10)$$

And then, by training a selector, all the weak classifiers have been trained from first to last and the best weak classifier (with the lowest estimated error) is determined accordingly [39]. The weak classifiers $H^{weak}$ correspondence to features. Consequently, a subset of $M$ features $F_{sub} = \{f_1, \cdots, f_m \mid f_i \in F\}$ can be selected from the global feature pool by the selectors.

#### 2) Principle of EOS-ELM

By using OS-ELM and online boosting algorithm as a weak classifier and an ensemble learning algorithm respectively, an EOS-ELM-based PRTSA model is presented. The specific steps of EOS-ELM training process are as follows:

(a) Initialize a group of $N$ selectors $h_1^{sel}, \cdots, h_N^{sel}$ randomly. Then, update all the selectors, when a new training sample $< x, t >$ is received. The weak classifier which has the smallest error will be chosen by the selector in the following way:

$$\arg\min_m(e_{n,m}), \quad e_{n,m} = \frac{\lambda_{n,m}^{wrong}}{\lambda_{n,m}^{wrong} + \lambda_{n,m}^{correct}} \quad (11)$$

where $e_{n,m}$ is the corresponding classification error rate of the classifier $h_{n,m}^{weak}$, which is the $m$-th weak classifier in the $n$-th selector; $\lambda_{n,m}^{correct}$ and $\lambda_{n,m}^{wrong}$ are respectively the sum of the importance weights of the samples which are correctly and wrongly classified at present.

(b) Update the importance $\lambda$ and the corresponding voting $\alpha_n$ of the sample, and pass them to the next selector $h_{n+1}^{sel}$. In this way, all the selectors repeat this procedure in turn.

(c) Finally, a strong classifier can be obtained by a linear combination of the corresponding weak classifiers selected by each selector, which is defined as

$$h^{strong}(x) = \text{sgn}\left(\sum_{n=1}^{N} \alpha_n \cdot h_n^{sel}(x)\right) \quad (12)$$

where $\text{sgn}(\cdot)$ is a sign function.

### B. EOS-ELM-based TSA Model

As above mentioned, the conventional framework of PRTSA consists of two closely related phases: off-line training and on-line application [15, 25], as shown in Fig. 1. In the offline training phase, the learning machine ($LM$) is trained by using the offline sample set $(\mathbf{X}, \mathbf{Y})$, and then the mapping relation $\mathbf{Y}=f(\mathbf{X})$ of the ideal model is obtained. In the online test phase, the transient stability assessment is executed on testing samples by using the trained model, and the transient stability status is predicted accordingly.

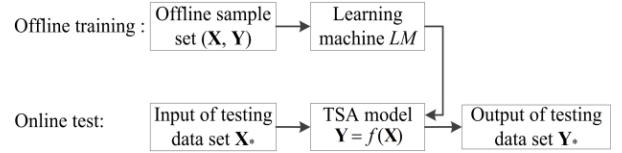

Fig. 1. Conventional framework of PRTSA

However, in the practical application, if a trained PRTSA model is unsatisfactory for some special samples in on-line application, the running model has to be terminated and offline retrained again, resulting in inefficiency for real-time implementation and lacking the on-line model updating ability [31, 35]. For solving this problem, an EOS-ELM-based TSA Model is proposed with shown in Fig. 2.

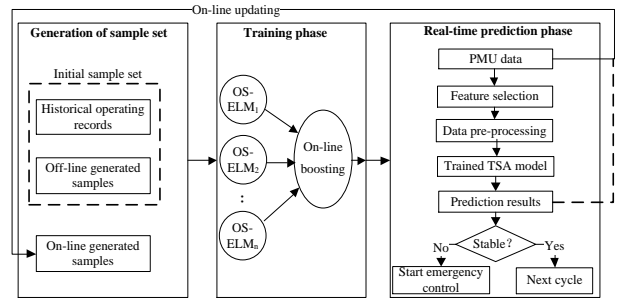

Fig. 2. EOS-ELM-based TSA model

#### 1) Generation of knowledge base

As is known, the generalization ability of a PRTSA model largely depends on the completeness and representativeness of the utilized knowledge base (KB) [21, 26]. For this reason, large amounts of time-domain simulations have been carried out to cover all of the typical contingencies as many as possible. Every operating point can be characterized by a stability index under contingencies and a vector of input features. By this means, the transient stability of a power system can be depicted by KB.

It is important to clarify that the generation scheme of KB can be available in off-line and on-line modes. Currently, the performance index of online dynamic security assessment (DSA) has met the requirements of practical application and become an important functional module of energy management systems (EMS) [40]. As far as the generation of on-line sample is concerned, the prospective operating points can be quickly generated by means of the very short-term load forecasting. Furthermore, parallel and distributed computation techniques are able to be used to greatly improve the efficiency of time-domain simulations [9]. In this way, large amounts of on-line samples can be generated from on-line simulation data reflecting the current operating modes through interfacing with the online DSA module of EMS.

*2) Operating modes*

The proposed approach has three modes of operation: the off-line learning mode, the on-line learning mode, and the real-time prediction mode.

**(1) Off-line learning mode**

The off-line learning model formulates the initial structure of the TSA model, which reflects the main transient characteristics of power systems [21, 25]. In this mode, training samples are extracted from the offline simulation data, which cover the combination of typical operating modes and contingencies. And then, the nonlinear relationship mapping between the system operation condition and the transient stability status of power systems is set up by training the model in the offline manner.

**(2) On-line learning mode**

In this mode, the proposed approach is able to extract samples from on-line simulation data reflecting the current operating modes through interfacing with the online DSA module of EMS. On the other hand, by learning new samples, the prediction model comprising its structure and parameters can be efficiently updated whenever a new special case occurs [35]. In this way, the present method is able to adapt the current operation modes of power systems; furthermore, the performance of the proposal is able to maintain accurate and more robust.

**(3) Real-time prediction mode**

In the proposal, it is supposed that once a large disturb occurs, this operation mode will be immediately triggered by a tripping signal issued by relay protection devices [19]. And then, the transient stability status will be predicted in real time according to the mapping relationships in the trained TSA model.

## IV. FEATURE SELECTION FOR EOS-ELM

### A. Construction of the Original Features

The used features in previous works are mainly pre-fault static features because the traditional supervisory control and data acquisition (SCADA) measurements are unable to provide wide-area post-fault dynamic information [17, 26]. Considering the matured application of WAMS, the proposal focuses on extracting input features from post-fault dynamic information besides static information to take full advantage of PMU data.

After having studied the literature comprehensively and carried out extensive simulations, a group of system-level classification features are constructed as the original feature set **A** [26], as listed in Table I. Here, $t_{cl+3c}$, $t_{cl+6c}$ and $t_{cl+9c}$ respectively are the 3rd, 6th and 9th cycle after the fault.

TABLE I
THE ORIGINAL INPUT FEATURES

| No. | Input features |
|---|---|
| Tz1 | Mean value of all the mechanical power before the fault incipient time |
| Tz2 | Maximum value of all the initial rotor acceleration rates at $t_0$ |
| Tz3 | Initial rotor angle of the machine with the maximum acceleration rate at $t_0$ |
| Tz4 | Mean value of all the initial acceleration power at $t_0$ |
| Tz5 | Value of system impact at $t_{cl}$ |
| Tz6 | Rotor angle of the machine with the biggest difference relative to the center of inertia at $t_{cl}$ |
| Tz7 | Kinetic energy of the machine with the maximum rotor angle at $t_{cl}$ |
| Tz8 | Rotor angle of the machine with the maximum kinetic energy at $t_{cl}$ |
| Tz9 | Maximum value of all the rotor kinetic energies at $t_{cl}$ |
| Tz10 | Mean value of all the rotor kinetic energies at $t_{cl}$ |
| Tz11 | Maximum value of the difference of rotor angles at $t_{cl}$ |
| Tz12 | Rotor angular velocity of the machine with the biggest difference relative to the center of inertia at $t_{cl}$ |
| Tz13 | Value of system impact at $t_{cl+3c}$ |
| Tz14 | Maximum value of all the rotor kinetic energies at $t_{cl+3c}$ |
| Tz15 | Mean value of all the rotor kinetic energies at $t_{cl+3c}$ |
| Tz16 | Rotor angle of the machine with the biggest difference relative to the center of inertia at $t_{cl+3c}$ |
| Tz17 | Maximum value of the difference of rotor angles at $t_{cl+3c}$ |
| Tz18 | Kinetic energy of the machine with the maximum rotor angle at $t_{cl+3c}$ |
| Tz19 | Rotor angular velocity of the machine with the biggest difference relative to the center of inertia at $t_{cl+3c}$ |
| Tz20 | Value of system impact at $t_{cl+6c}$ |
| Tz21 | Maximum value of all the rotor kinetic energies at $t_{cl+6c}$ |
| Tz22 | Mean value of all the rotor kinetic energies at $t_{cl+6c}$ |
| Tz23 | Kinetic energy of the machine with the maximum rotor angle at $t_{cl+6c}$ |
| Tz24 | Rotor angle of the machine with the biggest difference relative to the center of inertia at $t_{cl+6c}$ |
| Tz25 | Maximum value of the difference of rotor angles at $t_{cl+6c}$ |
| Tz26 | Rotor angular velocity of the machine with the biggest difference relative to the center of inertia at $t_{cl+6c}$ |
| Tz27 | Value of system impact at $t_{cl+9c}$ |
| Tz28 | Kinetic energy of the machine with the maximum rotor angle at $t_{cl+9c}$ |
| Tz29 | Maximum value of all the rotor kinetic energies at $t_{cl+9c}$ |
| Tz30 | Mean value of all the rotor kinetic energies at $t_{cl+9c}$ |
| Tz31 | Rotor angle of the machine with the biggest difference relative to the center of inertia at $t_{cl+9c}$ |
| Tz32 | Maximum value of the difference of rotor angles at $t_{cl+9c}$ |
| Tz33 | Rotor angular velocity of the machine with the biggest difference relative to the center of inertia at $t_{cl+9c}$ |

### B. BinJaya-based Feature Selection

In this work, a novel BinJaya algorithm with kernelized fuzzy rough sets (KFRS) is proposed for selecting an optimal feature subsets from the entire feature space constituted by a group of system-level classification features.

*1) Class Separability Criterion*

A classification task can be formulated as $<U, A, D>$, where $U$ is the nonempty and finite set of samples, $A$ is the set of features characterizing the classification, $D$ is the class attribute which divides the samples into subset $\{d_1, d_2, \cdots, d_m\}$.

Given $<U, A, D>$, a KFRS-based generalized classification function $gc(D)$ is used as the class separability criterion [26].

$$gc(D) = \left[ g\gamma_B^\theta(D) + g\omega_B^{\theta-\sigma}(D) \right] / 2 \quad (13)$$

where $B$ is the feature space $B \subseteq A$ and $B \neq \varnothing$, $g\gamma_B^\theta(D)$ and $g\omega_B^{\theta-\sigma}(D)$ are respectively the generalized dependency function and generalized classification certainty function.

### 2) Jaya algorithm

The 'Jaya' proposed by Rao in 2015 [41] is based on "get the victory by avoiding all failures" principle, and has been successfully used for solving engineering problems [42, 43].

Let $f(x)$ is the objective function to be minimized. At iteration $i$, assume that there are '$m$' number of design variables, '$n$' number of candidate solutions. Let the best (/worst) candidate $best$ (/$worst$) obtains the best (/worst) value of $f(x)$ in the entire candidate solutions. If $X_{j,k,i}$ is the value of the $j^{th}$ variable for the $k^{th}$ candidate during the $i^{th}$ iteration, then the value is modified as [41]:

$$X'_{j,k,i} = X_{j,k,i} + r_{1,j,i}(X_{j,best,i} - |X_{j,k,i}|) - r_{2,j,i}(X_{j,worst,i} - |X_{j,k,i}|) \quad (14)$$

where, $X_{j,best,i}$ (/$X_{j,worst,i}$) is the value of the variable $j$ for the $best$ (/$worst$) candidate. $X'_{j,k,i}$ is the updated value of $X_{j,k,i}$ and $r_{1,j,i}$ and $r_{2,j,i}$ are the two random numbers for the $j^{th}$ variable during the $i^{th}$ iteration in the range [0, 1]. $X'_{j,k,i}$ is accepted if it gives better function value. All the accepted function values at the end of iteration are maintained and these values become the input to the next iteration, shown as Fig. 3.

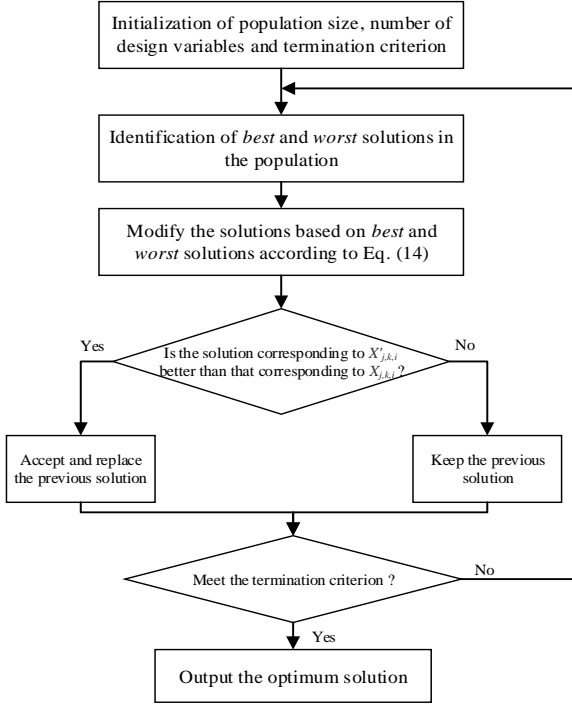

Fig. 3. Flowchart of the Jaya algorithm

### 3) Angle Modulation

The Jaya was originally developed for continuous-valued space. This paper employs angle modulation to enable the Jaya to correctly operate in binary space. The BinJaya is a Jaya algorithm that utilizes a trigonometric function as a bit string generator. Based on angle modulation, the function is derived from a signal processing technique. The technique uses a composed sin/cos generating function [44]:

$$g(y) = \sin(2\pi(y-o) \times p \times \cos(2\pi \times r(y-o))) + s \quad (15)$$

where $y$ is a single element from a set of evenly separated intervals determined by the number of bits specified to be generated. The coefficient $o$ represents the horizontal shift of the function, $p$ represents the maximum frequency of the sin function, $r$ represents the frequency of the cos function and $s$ represents the vertical shift of the function.

The standard Jaya is applied to optimize a simpler 4-dimensional tuple ($o, p, r, s$) representing the parameters of (15). After the iteration, the parameters are substituted back into (15). The resultant function is then sampled at the evenly spaced intervals to generate a bit for each interval. If the output value is positive, the bit value is noted as 1, else it is noted as 0 [44].

## V. CASE STUDY

In this section, the effectiveness of the proposal is examined using two testing cases: the IEEE 39-bus system and a real provincial system in China. All the simulations are executed under the MATLAB environment on a PC platform with 2 Intel Core dual core CPUs (2.4 GHz) and 6 GB RAM.

### A. Case 1—IEEE 39-bus System

First of all, the IEEE 39-bus system is used to test the proposal's effectiveness. The system (including 10 generators, 39 buses, 12 transformers and 34 lines) is a widely used testing case for examining the performance of a TSA approach [14-17], [19-21], [25, 26, 31, 35], and its single-line diagram is demonstrated in Fig. 4. The system represents a 345 kV power network in New England, USA.

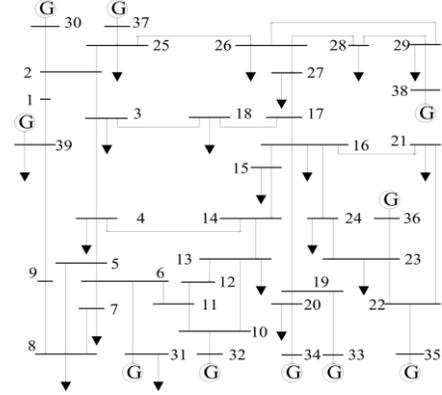

Fig. 4. IEEE 39-bus system

### 1) Generation of KB

In order to ensure KB with the adequate completeness and representativeness, large amounts of time-domain simulations have been executed [26]. The simulation calculation conditions of the modeled system are as follows. The generator model employed is the four-order model with the IEEE DC1 excitation system; the load model is the constant model. The considered contingencies are three-phase to ground short-circuit faults, the fault clearing time is supposed to 5 cycles for all of the contingencies (the faults are created at 0.2 s and cleared at 0.3 s), and a total of 60 different fault locations are taken into account. Here, it is assumed that the network topology is not changed when the faults are cleared [25, 26]. The contingencies are repeatedly performed at 11 levels (80%, 85%, ... , 130% of the base load), and 5 kinds of generator output under each load level are randomly assigned. Finally, a KB with total 3300 samples is obtained. In the KB, 2200 samples are chosen as the training set, and the rest are the testing set.

A class label $Class\_Lable$ of each sample is denoted by a transient stability index which is related to the relative rotor angle deviation during the transient period of a disturbed power system [19]. The label $Class\_Lable$ is determined as

$$Class\_Lable = \text{sgn}(360^o - |\Delta\delta|_{max}) \qquad (16)$$

where $|\cdot|$ is the absolute value function, and $\Delta\delta_{max}$ is of the maximum relative rotor angle deviation between generators in the period. By plotting the rotor angle swing curves of all the generators, a stable case and an unstable case are respectively demonstrated in Fig. 5 and 6.

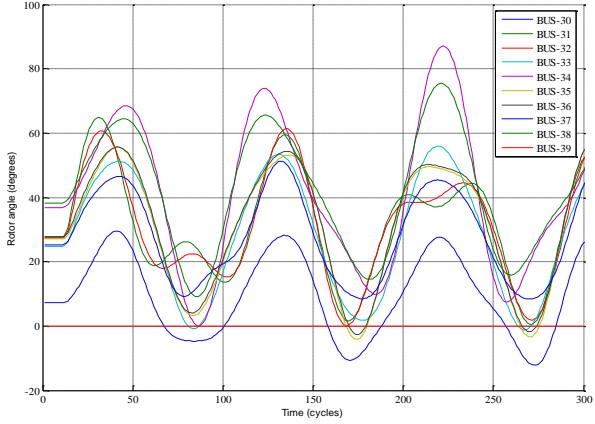

Fig. 5. Transient stable case

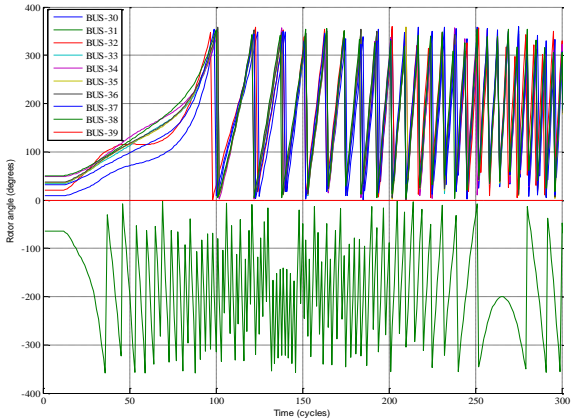

Fig. 6. Transient unstable case

### 2) Model selection

Note that, for OS-ELM, the sole parameter needed to be determined is the optimal number of hidden nodes $L$ [34, 35]. In this work, the parameter $L$ is determined by using the well-known cross-validation methods [26]. Specially speaking, the determination of the optimal network structure is implemented in such a way that the corresponding OS-ELM network offers the highest validation accuracy when the parameter $L$ achieves the optimal value. Among the common activation functions, the used one in this paper is the sigmoid function. This is selected because it gives the most satisfactory results when compared to other alternatives such as polynomials and RBF [34, 35].

Fig. 7 illustrates the validation results of OS-ELM.

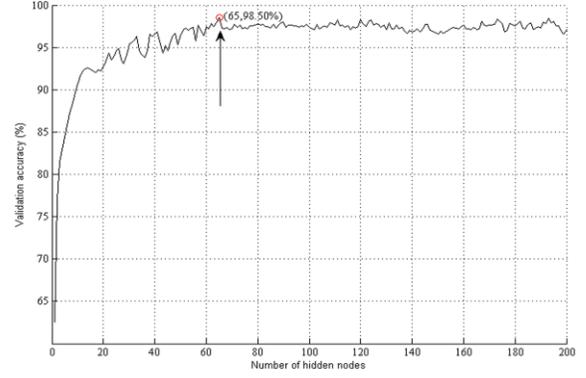

Fig. 7. Model selection of OS-ELM

In Fig. 7, the validation accuracy is plotted against the parameter $L$. Based on the given data set, it is noticeable to see that the validation accuracy can be dramatically improved with the increase of the parameter $L$ before reaching the maximum value, which is 98.50% with 65 hidden nodes. Consequently, the optimal value of $L$ is chosen to 65 in this testing case.

For EOS-ELM, the parameter $L$ in each weak classifier (an OS-ELM network) is assigned to the same value as the one used in the compared original OS-ELM network. The number of weak classifiers in EOS-ELM is in turn assigned to 5, 10, 15, 20 and 25 in 50 trials; then, the optimal number of weak classifiers is determined according to the standard deviation (SD) and the average testing accuracy of the 50 trials. More specifically, the optimal number of weak classifiers in EOS-ELM is chosen in such a way that the ensemble network is able to provide the better average testing accuracy and the lowest SD with the results obtained by OS-ELM for the same application. The results of EOS-ELM model selection for IEEE 39-bus system is shown in Table II.

TABLE II
MODEL SELECTION RESULTS OF EOS-ELM

| Number of networks | Average testing accuracy (%) | Testing SD |
|---|---|---|
| 5 | 98.62 | 0.0202 |
| **10** | **99.28** | **0.0097** |
| 15 | 98.88 | 0.0151 |
| 20 | 99.05 | 0.0186 |
| 25 | 98.91 | 0.0195 |

As observed from Table II, when the number of weak classifiers is chosen to 10, the corresponding EOS-ELM achieves the best predictive performance. On the one hand, the proposed model obtains the lowest SD value; on the other hand, the testing accuracy achieved by EOS-ELM is better than all the others as well. Therefore, the optimal number of weak classifiers in EOS-ELM is selected as 10 in our experiments.

### 3) Results and discussion

**(1) Comparison of EOS-ELM and original OS-ELM**

To examine the performance of the presented approach, a comparison of EOS-ELM and original OS-ELM in [35] is performed in one-by-one mode. By using the proposed BinJaya-based feature selection algorithm, the optimal feature subset $OFS_1$= {Tz4, Tz9, Tz19, Tz25, Tz26, Tz31, Tz32} can be obtained, and the test results are summarized in Table III.

Here, it should be noted that both the training time and accuracy are the average value of 50 trials of simulations in the table.

| Algorithms | Training time (s) | Accuracy | | SD | |
|---|---|---|---|---|---|
| | | Training (%) | Testing (%) | Training | Testing |
| OS-ELM | 0.0907 | 98.48 | 98.14 | 0.0036 | 0.0251 |
| **EOS-ELM** | **0.4531** | **99.85** | **99.28** | **0.0021** | **0.0097** |

The parameters are set as follows: for OS-ELM, the value of $L$ is set to 65, and the parameter $N_0$ used in the initialization phase is set to $N_0 = L + 50$; for EOS-ELM, the number of weak classifiers (OS-ELM networks) is 10, and the parameter $N_0$ for initialization phase is set to the same value as that of the compared OS-ELM.

As can be seen in Table III, the presented method outperforms original OS-ELM in almost all the performance indicators except for the training time. Compared with the original OS-ELM, the testing accuracy of EOS-ELM is increased by 1.14%, while at the same time the testing SD is decreased by 0.0154. This indicates that the classification accuracy and output stability of the original OS-ELM have been evidently strengthen through the use of ensemble learning. As a result, we can draw that ensemble learning is an effective way to improve the predictive performances of PRTSA models. Especially for the applications requiring high accuracy and reliability like TSA, EOS-ELM can just play its advantages in these respects.

**(2) Test results of other sequential learning algorithms**

In order to properly evaluate the effectiveness of the proposal, a test between the proposed approach and other popular sequential learning algorithms is further carried out. The performance of the proposal is compared with the algorithms, such as stochastic gradient descent back-propagation (SGBP) [45], Growing and Pruning Radial Basis Function (GAP-RBF) [46] and Minimal Resource Allocation Network (MRAN) [47], in one-by-one learning mode with the results summarized in Table IV. In Table IV, both the training time and the accuracy are the average values of 50 times.

| Algorithms | Training time (s) | Accuracy | | SD | |
|---|---|---|---|---|---|
| | | Training (%) | Testing (%) | Training | Testing |
| **EOS-ELM** | **0.4531** | **99.85** | **99.28** | **0.0021** | **0.0097** |
| SGBP | 0.0748 | 85.74 | 83.18 | 0.0155 | 0.0163 |
| GAP-RBF | 1.1525 | 95.42 | 93.82 | 0.0087 | 0.0251 |
| MRAN | 2.5162 | 96.66 | 95.06 | 0.0108 | 0.0322 |

The parameters used in this section are set in the following manner. For SGBP, the number of hidden neurons is set to 30, and the used activation function is the sigmoid additive activation function; the parameters of GAP-RBF and MRAN are fixed as: the distance parameters $\varepsilon_{max} = 0.5$, $\varepsilon_{min} = 0.01$, $\gamma = 0.99$, the impact factor adjustment parameter $\kappa = 0.80$.

As shown in Table IV, it can be concluded as follows:

(a) The performances of EOS-ELM, comprising the accuracy and stability, are far better than that of GAP-RBF and MRAN with much lower training time. The reason for this is that: for ELM, learning can be done without iterative tuning.

(b) Compared with SGBP, though the training time of EOS-ELM is a little more than that of SGBP, its testing accuracy is far superior to that of SGBP. In addition, the SD of EOS-ELM is less than that of SGBP, which suggests that the stability of EOS-ELM is better than that of SGBP. As can be seen, it is because the proposed method utilizes online ensemble learning that it has better stability and classification ability than SGBP.

Therefore, comprehensively considered with various related factors, EOS-ELM is the best method in this paper.

### B. Case 2—Real Power System of Liaoning Province

In order to further examine the applicability of the proposed method to a real system, the proposed approach is tested on the real power system of Liaoning province. The system is a large-scale power system in the northeast of China, which covers an area of 148,000 square kilometers. The total installed capacity of the system is about 39657.2 MW.

The modeled system contains 91 generators and 750 major buses in total. In addition, it has SVCs compensation and series compensated lines. The system has formed 5 connected channels with the external network through 10 500kV AC tie lines, $1 \pm 500$kV DC line and 1 500kV DC back-to-back converter station.

#### 1) Generation of KB

As same as in the Case-1, large amount of simulations have been executed. 12 out of all generators are modeled as the six-order model, and they are configured with the governors and the excitation systems; the rest generators are modeled as the classical machine model. The employed load model is the composite load model, which is made up of constant power load (60%) and constant-impedance load (40%).

The load level varies from 80% to 130% of the basic load. The fault type considered is the three-phase to ground fault, and the corresponding fault clearing times are set in the range from five to ten cycles. The locations of typical faults are set at different locations on lines (0, 25%, 50%, and 75% of the length). The stability criterion used here is consistent with that employed in Case-1. Finally, there are 2000 samples are totally created through time-domain simulations; 1320 of all the samples are randomly chosen to constitute the training set, and the rest as the testing set.

#### 2) Prediction results and performance

With the use of the presented feature selection scheme, the obtained optimal feature subset is $OFS_2 = \{Tz1, Tz4, Tz9, Tz17, Tz18, Tz19, Tz24, Tz25, Tz26, Tz31, Tz32, Tz33\}$. Moreover, by means of the model selection scheme in Case-1, the optimal number of OS-ELM networks in EOS-ELM is selected as 15 through large amounts of experiments.

In order to evaluate the prediction performance of the chosen optimal feature subset $OFS_2$ reasonably, it is used as the input for the proposed TSA model. At the same time, the $OFS_2$ has been compared and contrasted to the obtained $OFS_1$ in Case-1 and the original feature set **A** with the results shown in Table

V, where both the training and testing accuracies are the average values of 50 times, as in Case-1.

TABLE V
TEST RESULTS IN THE POWER SYSTEM OF LIAONING PROVINCE

| Algorithms | Feature set | Dimension | Accuracy | |
|---|---|---|---|---|
| | | | Training (%) | Testing (%) |
| EOS-ELM | OFS$_1$ | 7 | 96.06 | 95.88 |
| | **OFS$_2$** | **12** | **98.62** | **98.24** |
| | **A** | 33 | 98.65 | 98.20 |

Table V demonstrates that the proposed approach is able to predict the transient stability for the real power system. It can be observed that the classification performance of OFS$_2$ has similar classification performances with the original feature set **A**, while the dimension of the input space is sharply reduced to about one-third of its initial value (from 33 to 12).

Furthermore, it also illustrates that the prediction performance of OFS$_2$ is better than that of OFS$_1$. The reason for this is that, with the increase of the system size, the complexity of the stability pattern space of the disturbed system correspondingly increases [26], and thereby the predictive model needs more input features to more adequately represent the transient stability characteristics of the power system.

## VI. CONCLUSIONS

PRTSA has proved to be an effective way to determine the transient stability status of power systems. However, many of the existing PRTSA methods suffer from problems of inefficiency for real-time implementation and lacking the on-line model updating ability. To overcome this issue, a novel PRTSA approach based on EOS-ELM with BinJaya-based feature selection is proposed with the use of PMU data. The effectiveness of the proposal is examined, and the main conclusions are drawn from the simulation results as follows:

(1) The proposal has superior computation speed and prediction accuracy than other state-of-the-art sequential learning algorithms, including SGBP, GAP-RBF and MRAN.

(2) The proposed BinJaya algorithm can effectively solve the feature selection problem of PRTSA. Without sacrificing the classification performance, the dimension of the input space has been reduced to about one-third of its initial scale.

(3) The presented method can greatly improve the stability and generalization ability of an original OS-ELM with the use of ensemble learning techniques.

In future work, it is possible to use the proposal as a trigger for wide-area protection and control systems by predicting the impending power system transient instability. Furthermore, the BinJaya-based feature selection may be applied to any similar pattern classification problem in the area of engineering.